\documentclass[twocolumn,showpacs,preprintnumbers,amsmath,amssymb]{revtex4}

\usepackage{graphicx}
\usepackage{dcolumn}
\usepackage{bm}

\newcommand{\bq}{\begin{equation}}
\newcommand{\eq}{\end{equation}}
\newcommand{\bqa}{\begin{eqnarray}}
\newcommand{\eqa}{\end{eqnarray}}
\newcommand{\nn}{\nonumber \\}

\def\be     {\begin{equation}}
\def\ee     {\end{equation}}
\def\bea        {\begin{eqnarray}}
\def\eea        {\end{eqnarray}}
\def\bnn    {\begin{eqnarray*}}
\def\enn    {\end{eqnarray*}}

\begin{document}

\title{Effect of nonmagnetic disorder on criticality in the "dirty" U(1) spin liquid}
\author{Ki-Seok Kim}
\affiliation{Korea Institute for Advanced Study, Seoul 130-012,
Korea}
\date{\today}

\begin{abstract}
We investigate the effect of nonmagnetic disorder on the stability
of the algebraic spin liquid ($ASL$) by deriving an effective
field theory, nonlinear $\sigma$ model ($NL{\sigma}M$). We find
that the anomalous critical exponent characterizing the
criticality of the $ASL$ causes an anomalous gradient in the
$NL{\sigma}M$. We show that the sign of the anomalous gradient
exponent or the critical exponent of the $ASL$ determines the
stability of the "dirty" $ASL$. A positive exponent results in an
unstable fixed point separating delocalized and localized phases,
which is consistent with our previous study [Phys. Rev. B {\bf
70}, 140405 (2004)]. We find power law suppression for the density
of spinon states in contrast to the logarithmic correction in the
free Dirac theory. On the other hand, a negative exponent
destabilizes the $ASL$, causing the Anderson localization. We
discuss the implication of our study in the pseudogap phase of
high $T_c$ cuprates.
\end{abstract}


\maketitle

The effect of nonmagnetic disorder in the presence of strong
correlations between electrons is one of the central interests in
modern condensed matter physics. This situation is realized when
nonmagnetic impurities are doped in the Mott insulator. For
example, $Zn$ doped high $T_c$ cuprates are those. In order to
solve this problem it should be clarified how to describe the Mott
insulator. Here we concentrate on a paramagnetic Mott insulator
($PMMI$). According to one possible scenario, hole doping to a
$PMMI$ results in high $T_c$
superconductivity\cite{Doped_PMMI_SC}. Understanding the nature of
the $PMMI$ may be crucial for the mechanism of high $T_c$
superconductivity. Recently, the $PMMI$ was proposed to be the
U(1) spin liquid ($U1SL$)\cite{U1SL}. The $U1SL$ is the state
described by $QED_3$ in terms of massless Dirac spinons strongly
interacting via noncompact U(1) gauge fields\cite{U1SL}. The key
feature of the $U1SL$ lies in the criticality characterized by
anomalous critical exponents originating from long range gauge
interactions\cite{Wen_ARPES,Wen_spin_correlation,DonKim_QED,Tesanovic_QED,
AFL_spin_correlation,Ye_propagator,Khveshchenko_propagator,QED_eta}.
All correlation functions show algebraic power law decay with
anomalous critical exponents, resulting in no well defined
quasiparticles (spinons)\cite{Wen_ARPES,Wen_spin_correlation,
DonKim_QED,Tesanovic_QED,AFL_spin_correlation,Ye_propagator,
Khveshchenko_propagator,QED_eta}. Owing to the power law behavior
the $U1SL$ is sometimes called the algebraic spin liquid ($ASL$)
more appropriately.

Recently, the role of nonmagnetic impurities in the $ASL$ was
investigated by the present author\cite{Disorder_ASL}. In contrast
to the free Dirac theory in two spacial
dimensions\cite{Lee_disorder,Fisher_disorder} long range gauge
interactions can induce a delocalized state at zero
temperature\cite{Disorder_ASL}. The presence of nonmagnetic
disorder destabilizes the free Dirac fixed point. The
renormalization group ($RG$) flow goes away from the fixed point,
indicating localization\cite{Lee_disorder,Fisher_disorder}. On the
other hand, the $ASL$ fixed point in $QED_3$\cite{U1SL} remains
stable at least against weak disorder\cite{Disorder_ASL}. An
unstable fixed point separating delocalized and localized phases
is found\cite{Disorder_ASL}. The $RG$ flow shows that the effect
of disorder potential vanishes if we start from sufficiently weak
disorder.

In the present study we reexamine the effect of nonmagnetic
disorder on the $ASL$ by deriving an effective field theory,
nonlinear $\sigma$ model ($NL{\sigma}M$). The main idea is firstly
to integrate out gauge fluctuations. This gives the single
particle propagator Eq. (2) with the anomalous exponent
$\eta_{\psi}$, the hall mark of the $ASL$. Based on this $ASL$
propagator we obtain the $NL{\sigma}M$ Eq. (8) by usual treatment.
The core in the $NL{\sigma}M$ is the anomalous gradient exponent
$\eta_{\sigma}$ originating from the critical exponent
$\eta_{\psi}$ of the $ASL$. We find that the sign of
$\eta_{\sigma}$ or $\eta_{\psi}$ determines the stability of the
$ASL$ against nonmagnetic impurities. In the case of
$\eta_{\sigma} > 0$ an unstable fixed point separating delocalized
and localized phases is found, which is consistent with our
previous study\cite{Disorder_ASL}. The density of spinon states
and the conductance of a spinon number (internal gauge charge)
show power law corrections in contrast to the logarithmic
suppression in the free Dirac theory\cite{Fisher_disorder}. In the
case of $\eta_{\sigma} < 0$ the $ASL$ becomes unstable. The
Anderson localization is expected. We discuss the implication of
our study in the pseudogap phase of high $T_c$ cuprates.

First we briefly review how the effective $QED_3$ Lagrangian is
derived from the antiferromagnetic Heisenberg model, $H =
J\sum_{<i,j>}\vec{S}_{i}\cdot\vec{S}_{j}$ with $J>0$. Inserting
the fermionic representation of spin $\vec{S}_{i} =
\frac{1}{2}f^{\dagger}_{i\alpha}\vec{\tau}_{\alpha\beta}f_{i\beta}$
into the Heisenberg model, and performing the standard
Hubbard-Stratonovich transformation for an exchange hopping
channel, we obtain an effective one body Hamiltonian for the
fermions coupled to an order parameter, $H_{eff} =
-J\sum_{<i,j>}f_{i\alpha}^{\dagger}\chi_{ij}f_{j\alpha} - h.c.$.
Here $f_{i\alpha}$ is a fermionic spinon with spin $\alpha =
\uparrow, \downarrow$, and $\chi_{ij}$ is an auxiliary field
called a hopping order parameter. Notice that the hopping order
parameter $\chi_{ij}$ is a complex number defined on links $ij$.
Thus, it can be decomposed into $\chi_{ij} =
|\chi_{ij}|e^{i\theta_{ij}}$ where $|\chi_{ij}|$ and $\theta_{ij}$
are the amplitude and phase of the hopping order parameter,
respectively. Inserting this representation of $\chi_{ij}$ into
the effective Hamiltonian, we obtain $H_{eff} =
-J\sum_{<i,j>}|\chi_{ij}|f_{i\alpha}^{\dagger}e^{i\theta_{ij}}f_{j\alpha}
- h.c.$. Then, we can easily see that this effective Hamiltonian
has internal U(1) gauge symmetry
$H'_{eff}[f'_{i\alpha},\theta'_{ij}] =
H_{eff}[f_{i\alpha},\theta_{ij}]$ under the following U(1) phase
transformations, $f'_{i\alpha} = e^{i\phi_{i}}f_{i\alpha}$ and
$\theta'_{ij} = \theta_{ij} + \phi_{i} - \phi_{j}$. This implies
that the phase $\theta_{ij}$ of the hopping order parameter plays
the same role as the U(1) gauge field $a_{ij}$. When a spinon hops
on lattices, it obtains the Aharnov-Bohm phase owing to the U(1)
gauge field $a_{ij}$. It is well known that the stable mean field
phase is $\pi$ flux state at half filling\cite{DonKim_QED}. This
means that a spinon gains the phase of $\pi$ when it turns around
one plaquette. In the $\pi$ flux phase low energy elementary
excitations are massless Dirac spinons near nodal points showing
gapless Dirac spectrum and U(1) gauge
fluctuations\cite{DonKim_QED}. In the low energy limit the
amplitude $|\chi_{ij}|$ is frozen to be $|\chi_{ij}| =
J|<f_{j\alpha}^{\dagger}f_{i\alpha}>| \equiv \chi_{0}$. As a
result we obtain the following low energy effective Lagrangian in
terms of massless Dirac fermions near the nodal points interacting
via compact U(1) gauge fields\cite{U1SL} \bqa && Z =
\int{D\psi_{n\sigma}}{Da_{\mu}}e^{-\int{d^3x} {\cal L}} , \nn &&
{\cal L} = \sum_{\sigma=\uparrow,\downarrow}\sum_{n=1}^{2}
\bar{\psi}_{n\sigma}\gamma_{\mu}(\partial_{\mu} +
ia_{\mu})\psi_{n\sigma} + \frac{1}{2e^2}|\partial\times{a}|^2 .
\eqa Here $\psi_{n\sigma}$ is the two component massless Dirac
spinon, where $n = 1, 2$ represent the nodal points,
$(\pi/2,\pi/2)$ and $(\pi/2,-\pi/2)$, and ${\sigma} = \uparrow,
\downarrow$, SU(2) spin. They are given by $\psi_{1\sigma} =
\left( \begin{array}{c} f_{1e\sigma} \\ f_{1o\sigma} \end{array}
\right)$ and $\psi_{2\sigma} = \left( \begin{array}{c} f_{2o\sigma} \\
f_{2e\sigma} \end{array} \right)$, respectively. In the spinon
field $f_{nl\sigma}$ $n = 1, 2$ represent the nodal points, $l =
e, o$, even and odd sites, and $\sigma = \uparrow, \downarrow$,
its spin, respectively\cite{DonKim_QED}. The Dirac matrices
$\gamma_{\mu}$ are given by the Pauli matrices $\gamma_{\mu} =
(\sigma_{3}, \sigma_{2}, \sigma_{1})$ where they satisfy the
Clifford algebra $[\gamma_{\mu},\gamma_{\nu}]_{+} =
2\delta_{\mu\nu}$\cite{DonKim_QED}. $a_{\mu}$ is the U(1) gauge
field. The kinetic energy of the gauge field results from
particle-hole excitations of high energy spinons. $e$ is an
effective internal charge, not a real electric charge.

We have difficulty in solving Eq. (1) owing to instanton
excitations originating from the compactness of the U(1) gauge
field. Instantons represent tunnelling events between
energetically (nearly) degenerate but topologically inequivalent
vacua. In the U(1) gauge theory the instanton has a magnetic
monopole configuration. Magnetic monopoles are known to play a
crucial role in confinement physics\cite{Polyakov,U1SL}. Recently,
it was shown that the instanton effect can be suppressed at least
in the large $N$ limit ($\sigma = \uparrow, \downarrow$
$\rightarrow$ $\sigma = 1, 2, ..., N$) where $N$ is the flavor
number of massless Dirac fermions\cite{U1SL}. Existence of quantum
criticality in the $QED_3$ with noncompact U(1) gauge fields is
the key mechanism of suppression of instanton excitations. It is
well known that the $QED_3$ has a stable charged fixed point in
the large $N$ limit. Hermele et. al examined the stability of the
charged fixed point against instanton excitations\cite{U1SL}. At
the fixed point charges can be sufficiently screened by critical
fluctuations of Dirac fermions. The larger the flavors $N$ of the
Dirac fermions, the smaller electric charges, thus resulting in
huge magnetic charges. This can cause a negative scaling dimension
to instanton fugacity\cite{U1SL}. As a result instanton
excitations can be suppressed at the charged fixed point in the
large $N$ limit. Deconfinement of charged matter fields, here the
massless Dirac spinons can be achieved at the quantum critical
point. The $QED_3$ with noncompact gauge fields appears as the
critical field theory at the charged fixed point. All correlation
functions exhibit power law behaviors with anomalous critical
exponents generated from long range gauge interactions. In this
respect the state described by the $QED_3$ is called the $ASL$ and
the charged fixed point, the $ASL$ critical point.

As discussed in the above, the quantum criticality of the
effective $QED_3$ Eq. (1) is the crucial point for deconfinement
of the Dirac spinons. But, we should remember that the $ASL$
criticality can survive only in the limit of large flavors $N$
corresponding to the SU(N) quantum antiferromagnet ($\sigma = 1,
2, ..., N$). In the case of the physical SU(2) antiferromagnet
($\sigma = \uparrow, \downarrow$) it is not clear if the $ASL$
criticality remains owing to spontaneous chiral symmetry breaking
($S\chi{S}B$) causing antiferromagnetism. It is believed that
there exists the critical flavor number $N_c$ associated with the
$S\chi{S}B$ in the
$QED_3$\cite{CSB1,CSB2,DonKim_QED,Tesanovic_QED,U1SL}. But, the
precise value of the critical number is far from
consensus\cite{CSB2}. If the critical value is larger than $2$,
the $S\chi{S}B$ is expected to occur for the physical $N = 2$
case. Then, the Dirac fermions become massive. In this case the
massive Dirac spinons are confined to form spin $1$ excitations,
antiferromagnons\cite{DonKim_QED}. On the other hand, in the case
of $N_c < 2$ the $ASL$ criticality remains stable against the
$S\chi{S}B$. In the present paper we assume that the chiral
symmetry is preserved even in the SU(2) quantum antiferromagnet.
There is a supporting argument for existence of the $ASL$ even in
the case of $N_c > 2$\cite{Doped_PMMI_SC}. If there are additional
massless fluctuations carrying internal U(1) gauge charges, the
$S\chi{S}B$ can be forbidden. These additional gapless
fluctuations can arise from quantum critical points in association
with antiferromagnetism or superconductivity. For example,
critical fluctuations of doped holes near the superconducting
transition quantum critical point are expected to increase the
flavor number of massless fluctuations\cite{Nc}. If the total
flavor number of massless Dirac spinons and doped holes exceeds
the critical value $N_c$, the $S\chi{S}B$ is not expected to
occur\cite{Nc,Critical_holon}. In this case the quantum critical
point would be described by the $ASL$ for spin degrees of
freedom\cite{Doped_PMMI_SC}. The role of doped holes in the $ASL$
will be discussed later in more detail.

The criticality of the $ASL$ is characterized by the critical
exponents of correlation functions. The single particle propagator
$G_{ASL}(k) = <\psi_{n\sigma}(k)\bar{\psi}_{n\sigma}(k)>$ is given
by\cite{Wen_ARPES,Tesanovic_QED,Ye_propagator,Khveshchenko_propagator,QED_eta}
\bqa G_{ASL}(k) \approx
-i\frac{\gamma_{\mu}k_{\mu}}{k^{2-\eta_{\psi}}} , \eqa where
$\eta_{\psi}$ is the anomalous critical exponent of the $ASL$
propagator. Here we briefly sketch how the $ASL$ propagator Eq.
(2) is derived from the effective $QED_3$ Eq. (1). The single
particle propagator is generally given by $G^{-1}_{ASL}(k) =
G_{0}^{-1}(k) + \Sigma(k)$, where $G_{0}^{-1}(k) =
i\gamma_{\mu}k_{\mu}$ is the inverse of the bare spinon
propagator, and $\Sigma(k)$, the spinon self-energy resulting from
long range gauge interactions. In the usual $1/N$
expansion\cite{Wen_ARPES,Tesanovic_QED,DonKim_QED} the self-energy
is represented by $\Sigma(k) =
\int\frac{d^3q}{(2\pi)^{3}}\gamma_{\mu}G_{0}(k+q)\gamma_{\nu}D_{\mu\nu}(q)$,
where $D_{\mu\nu}(q)$ is the renormalized propagator of the U(1)
gauge field due to particle-hole excitations (polarization) of
massless Dirac fermions. The gauge propagator is obtained to be
$D_{\mu\nu}(q) \approx \Pi_{\mu\nu}^{-1}(q) =
\frac{8}{Nq}\Bigl(\delta_{\mu\nu} -
\frac{q_{\mu}q_{\nu}}{q^2}\Bigr)$ in the Lorentz gauge, where
$\Pi_{\mu\nu}(q) =
N\int\frac{d^3k}{(2\pi)^{3}}Tr[G_{0}(k)\gamma_{\mu}G_{0}(k+q)\gamma_{\nu}]$
is the polarization function of the Dirac fermions. Inserting this
gauge propagator into the expression of the self-energy, we obtain
the spinon self-energy of logarithmic momentum dependence
$\Sigma(k) =
i\eta_{\psi}\gamma_{\mu}k_{\mu}ln\Bigl(\frac{\Lambda}{k}\Bigr)$,
where $\eta_{\psi}$ is the anomalous exponent and $\Lambda$, the
momentum cutoff. The absolute value of the exponent $\eta_{\psi}$
is proportional to the inverse of the flavor number, i.e.,
$|\eta_{\psi}| \sim N^{-1}$\cite{Wen_ARPES,Tesanovic_QED,
Ye_propagator,Khveshchenko_propagator,QED_eta}. Since the $QED_3$
is the critical field theory at the charged critical point, all
correlators should exhibit power law behaviors. In this respect
the logarithmic momentum dependence should be considered to be the
lowest order in the power law behavior. As a result we can obtain
the $ASL$ single particle propagator Eq. (2) from the following
nonperturbative consideration\cite{Tesanovic_QED} $G_{ASL}^{-1}(k)
= i\gamma_{\mu}k_{\mu}\Bigl[1 +
\eta_{\psi}ln\Bigl(\frac{\Lambda}{k}\Bigr)\Bigr] \approx
i\gamma_{\mu}k_{\mu}\Bigl(\frac{\Lambda}{k}\Bigr)^{\eta_{\psi}}$.
Although the spinon propagator exhibits its algebraic form, it is
difficult to give a definite physical meaning. This is because it
is not gauge invariant. All physical observables should be gauge
invariant. The critical exponent $\eta_{\psi}$ should be evaluated
in a gauge invariant way. The following gauge invariant green
function is usually considered, $G_{ASL}(x) =
<T_{\tau}[\psi_{n\sigma}(x)e^{i\int_{0}^{x}d\zeta_{\mu}a_{\mu}(\zeta)}\bar{\psi}_{n\sigma}(0)]>$.
It is not easy to obtain the critical exponent $\eta_{\psi}$ by
calculating this gauge invariant green function. Its precise value
is far from consensus and under current debate. The crucial point
in the following discussion is the sign of the exponent
$\eta_{\psi}$. Most evaluations\cite{Wen_ARPES,
Ye_propagator,Khveshchenko_propagator,QED_eta} suggest its
negative sign, $\eta_{\psi} < 0$. However, as argued in Ref.
\cite{Tesanovic_QED}, its negative sign is unphysical in the sense
that the $ASL$ propagator is more coherent at long distances than
the propagator of the free Dirac theory owing to long range gauge
interactions. This result is completely in contrast to the usual
role of interactions. Interactions would make the propagator less
coherent. This is indeed true in the critical field theories with
local repulsive interactions (for example, $\phi^4$ theory).
Positive critical exponents are well known in these theories. If
the critical exponent $\eta_{\psi}$ is positive, the long range
gauge interactions destabilize the Fermi liquid pole (the pole of
the single particle green function in the free Dirac theory). The
renormalization factor $Z(p)$ representing weight of
quasiparticles is given by $Z(p) \sim p^{\eta_{\psi}}$ with
momentum $p$. In the case of $\eta_{\psi} > 0$ it vanishes in the
long wave length and low energy limit, i.e., $Z(p\rightarrow 0)
\rightarrow 0$, leading to Luttinger liquid-like power law
correlators. This is the $ASL$ as a critical state. In this
respect the $ASL$ can be considered to be two dimensional
realization of the one dimensional Luttinger
liquid\cite{Tesanovic_QED}. In the present paper we do not
determine its sign. Instead we use the exponent $\eta_{\psi}$ as a
phenomenological parameter. We consider both cases, $\eta_\psi <
0$ and $\eta_\psi > 0$. Here we assume that the $ASL$ green
function Eq. (2) is obtained in a gauge invariant
way\cite{Wen_ARPES,Tesanovic_QED,Ye_propagator,Khveshchenko_propagator,QED_eta}
and thus the critical exponent $\eta_{\psi}$ is gauge invariant.
We repeat that its absolute value is given by $|\eta_{\psi}| \sim
N^{-1}$ in the $1/N$ approximation\cite{Wen_ARPES,Tesanovic_QED,
Ye_propagator,Khveshchenko_propagator,QED_eta}.

The $ASL$ propagator $G_{ASL}(k) =
<\psi_{n\sigma}(k)\bar{\psi}_{n\sigma}(k)> =
-i\frac{\gamma_{\mu}k_{\mu}}{k^{2-\eta_{\psi}}}$ can be easily
obtained from the following Lagrangian \bqa && {\cal L}_{ASL} =
\sum_{\sigma=\uparrow,\downarrow}\sum_{n=1}^{2}
\bar{\psi}_{n\sigma}\gamma_{\mu}\partial_{\mu}\partial^{-\eta_{\psi}}\psi_{n\sigma}
. \eqa Integration over U(1) gauge fluctuations in the $QED_3$ Eq.
(1) results in the $ASL$ green function Eq. (2) as explicitly
demonstrated above. Eq. (2) is easily derived from Eq. (3).
Therefore, Eq. (3) can be considered to be an effective field
theory resulting from integration over the U(1) gauge field in the
$QED_3$ Eq. (1). In this respect the Dirac spinon field
$\psi_{n\sigma}$ in Eq. (3) is clearly different from that in Eq.
(1). It should be considered to be a renormalized field resulting
from the gauge interactions. The renormalized Dirac spinon field
arises from the self-energy correction. We view Eq. (3) as the
prototype describing the $ASL$. More generally, one may think that
Eq. (3) represents one class of critical field theories depending
on the critical exponent $\eta_{\psi}$. This point of view is
parallel to the standpoint that a free fermion theory is the
foundation describing Fermi liquid. One cautious theorist can
argue that the $ASL$ Lagrangian Eq. (3) is not sufficient because
Eq. (3) does not include appropriate vertex corrections. But this
guess is not correct. Remember that the critical exponent
$\eta_{\psi}$ was obtained in a gauge invariant way although the
calculation of $\eta_{\psi}$ in a gauge invariant way was not
explicitly demonstrated in the above owing to its
complexity\cite{Wen_ARPES,Tesanovic_QED,Ye_propagator,Khveshchenko_propagator,QED_eta}.
It is well known that we cannot satisfy the gauge invariance
without vertex corrections\cite{AFL_spin_correlation}. Thus, the
evaluation of $\eta_{\psi}$ in a gauge invariant way includes
vertex corrections. This argument can be checked by considering a
two particle green function. The critical exponent of a spin-spin
correlation function, obtained in a gauge invariant way including
vertex corrections explicitly in Eq. (1), is given by twice the
exponent of the single particle propagator in Eq. (3), i.e.,
$2\eta_{\psi}$ in the case of $\eta_{\psi} <
0$\cite{Wen_spin_correlation,AFL_spin_correlation}. This was also
pointed out in Ref. \cite{QED_eta}. This implies that we can treat
the $ASL$ effective Lagrangian Eq. (3) as a "free" theory.

We introduce a random potential $V({\bf r})$ in the $ASL$ [Eq.
(3)] \bqa && S = \int{d^{D}{x}}\sum_{\sigma =
\uparrow,\downarrow}\sum_{n=1}^{2}\Bigl[\bar{\psi}_{n\sigma}\gamma_{\mu}\partial_{\mu}
\partial^{-\eta_{\psi}}\psi_{n\sigma} +
V({\bf r})\bar{\psi}_{n\sigma}\psi_{n\sigma} \Bigr] . \eqa As the
randomness is independent of time and the ${\cal L}_{ASL}$ is
quadratic in $\psi_{n\sigma}$, each frequency sector in Eq. (4) is
separated from each other. Thus it is sufficient to consider only
zero frequency sector. This motivates us to explore $D=2$. Here we
assume that $V({\bf r})$ is a gaussian random potential of
$<V({\bf r})V({\bf r'})> = W\delta({\bf r} - {\bf r'})$ with
$<V({\bf r})> = 0$. Using the standard replica trick to average
over the gaussian random potential, we obtain the following
effective action in the presence of nonmagnetic disorder \bqa && S
= \int{d^{2}{\bf
r}}\sum_{\alpha=1}^{M}\Bigl(\bar{\Psi}_{\alpha}\Gamma_{i}\partial_{i}
\partial^{-\eta_{\psi}}\Psi_{\alpha} + i\epsilon\bar{\Psi}_{\alpha}\Psi_{\alpha} \Bigr)
\nn && - \frac{W}{2}\sum_{\alpha,\alpha' = 1}^{M}\int{d^{2}{\bf
r}}\bar{\Psi}_{\alpha}\Psi_{\alpha}\bar{\Psi}_{\alpha'}\Psi_{\alpha'}
. \eqa Here $\Psi_{\alpha} = \left(
\begin{array}{c}
\psi_{1\uparrow\alpha} \\ \psi_{2\uparrow\alpha} \\ \psi_{1\downarrow\alpha} \\
\psi_{2\downarrow\alpha} \end{array} \right)$ is the eight
component spinor where $\alpha = 1, ..., M$ is an replica index
and the limit $M \rightarrow 0$ is to be taken at the end.
$\Gamma_{\mu} = I \otimes \sigma^{3} \otimes \gamma_{\mu}$ is the
eight-by-eight gamma matrix. $I$ acts on SU(2) spin space and
$\sigma^3$, different nodal points. We have included an
infinitesimal imaginary potential $\epsilon$ in order to generate
correlation functions\cite{Fisher_disorder}. In Eq. (5) one can
easily read the bare scaling dimension of the disorder strength
$W$. It is given by $dim[W] = -2\eta_{\psi}$. In the case of
$\eta_\psi > 0$ this is consistent with our previous $RG$
study\cite{Disorder_ASL}. In the study the $RG$ equation is found
to be ${dW}/{dlnL} \approx - \chi{e}^{2}W$ to the first order in
$W$. $\chi$ is a positive numerical constant. Considering the
charged fixed point $e_{c}^{2} \sim N^{-1}$ with the flavor $N$,
the $RG$ equation yields $dim[W] = - \chi{e}_{c}^{2} \sim - N^{-1}
\sim - \eta_{\psi}$. But, in the opposite case of $\eta_\psi < 0$
the scaling dimension of the disorder strength $W$ is positive,
indicating instability of $W = 0$.

Performing the standard Hubbard-Stratonovich transformation and
the gaussian integration for the spinon field $\Psi_{\alpha}$, we
obtain the following effective action in terms of the order
parameter field ${\bf Q}$ \bqa && S_{eff} = \int{d^{2}{\bf
r}}\Bigl[ - \frac{1}{2W}Tr[{\bf Q}^2({\bf r})] \nn && +
Trln\Bigl(\Gamma_{i}\partial_{i}
\partial^{-\eta_{\psi}} + i{\bf Q} + i\epsilon\Bigr)
\Bigr]   . \eqa Here ${\bf Q}$ is the $8M\times{8}M$ matrix field.
Its saddle point is given by ${\bf Q}_{\alpha\alpha'} =
W<\bar{\Psi}_{\alpha}\Psi_{\alpha'}>$. The saddle point equation
is obtained to be   \bqa && 1 = 8W\int\frac{d^2{\bf
k}}{(2\pi)^{2}} \frac{1}{|{\bf k}|^{2-2\eta_{\psi}} + Q^2} \equiv
{W}F(Q^2) . \eqa Here we replaced ${\bf Q}_{\alpha\alpha'}$ with
${\bf Q}_{\alpha\alpha'} = \delta_{\alpha\alpha'}Q$. Since
$F(Q^2)$ is a monotonically decreasing function from infinity to
zero in the case of $\eta_{\psi} < 1$, there exists only one
nonzero solution for $Q$\cite{Saddle_point}. The relation of
$|\eta_\psi| \sim N^{-1}$ guarantees this consideration. In the
free Dirac theory ($\eta_{\psi} = 0$) $Q = \Lambda{e}^{-\pi/4W}$
with the momentum cutoff $\Lambda$ is obtained. In the $ASL$ the
meaning of nonzero $Q$ is not clear. In Fermi liquid it
corresponds to the imaginary part of the self energy and thus the
finite value of $Q$ generates a finite density of quasiparticle
states due to disorder. But, in the present case there are no well
defined quasiparticles. Thus we cannot define the density of
quasiparticle states. Mathematically, it gives an imaginary part
to the single particle propagator. In this respect we can think
that a diffusive behavior of critical spinons emerges by disorder.

Next we consider small fluctuations around this saddle point. Here
we follow Ref. \cite{Sigma_model}. Inserting the expression ${\bf
Q}_{\alpha\alpha'} = Q{\bf U}_{\alpha\alpha'}$ into Eq. (6) and
expanding the logarithmic term to the second order in ${\bf
U}_{\alpha\alpha'}$, we obtain the $NL{\sigma}M$\cite{Sigma_model}
\bqa && S_{NL\sigma{M}} = \int{d^2{\bf r}}\Bigl(
\frac{1}{2g_{\sigma}}Tr[\nabla{\bf U}^{\dagger}({\bf
r})\frac{1}{|\nabla|^{\eta_{\sigma}}}\nabla{\bf U}({\bf r})] \nn
&& + \epsilon{Tr}[{\bf U}({\bf r}) + {\bf U}^{\dagger}({\bf r})]
\Bigr) . \eqa Here $g_{\sigma}^{-1}$ is considered to be the
conductance $\sigma_{sn}$ of a spinon number (internal gauge
charge). Its bare value ${g_{\sigma}^{0}}^{-1}$ is obtained to be
${g_{\sigma}^{0}}^{-1} =
\frac{1}{2}\int\frac{d^2k}{(2\pi)^{2}}\Bigl[\frac{Q}{k^{2-2\eta_{\psi}}+Q^2}\Bigr]^{2}$.
The anomalous gradient exponent $\eta_{\sigma}$ is given by
$\eta_{\sigma} = 2\eta_{\psi}$. As mentioned in the introduction,
the anomalous gradient is the crucial feature in the
$NL{\sigma}M$. Its existence can be interpreted to be the mirror
of the criticality $\eta_{\psi}$ in the original system, here the
$ASL$. Generally speaking, we are considering the role of
nonmagnetic disorder in the critical systems with anomalous
critical exponents. The stability of the $ASL$ critical point is
determined by the anomalous gradient exponent $\eta_{\sigma}$ or
the critical exponent $\eta_{\psi}$ of the $ASL$. In the case of
$\eta_\sigma > 0$ ($\eta_\psi > 0$) the $ASL$ critical point
remains stable against weak randomness. But in the case of
$\eta_\sigma < 0$ ($\eta_\psi < 0$) it becomes unstable and the
Anderson localization occurs. We discuss this main issue by
investigating the $RG$ equation of the stiffness parameter
$g_{\sigma}$.

One can easily derive the following $RG$ equation in one loop
order\cite{Polyakov} \bqa && \frac{dg_{\sigma}}{dlnL} = -(d-2 +
\eta_{\sigma})g_{\sigma} + A{g}_{\sigma}^{2} . \eqa Here $d$ is
the spacial dimension and $d = 2$ is the present case. $A$ is a
positive numerical constant. In the case of $\eta_{\psi}
\rightarrow 0$ the above $RG$ equation is reduced to that of the
free Dirac theory\cite{Fisher_disorder}. Furthermore, this
equation corresponds to Eq. (7) in our previous
study\cite{Disorder_ASL}. Inserting $g_{\sigma} \rightarrow W$ and
$\eta_{\sigma} \sim N^{-1} \rightarrow e^{2}$ into Eq. (9), we
obtain the following $RG$ equation of the disorder strength $W$,
${dW}/{dlnL} = - {e}^{2}W + AW^{2}$. This completely coincides
with Eq. (7) in the previous study\cite{Disorder_ASL}.

Now we discuss the phases from the $RG$ equation (9). First we
consider the case of a positive exponent, $\eta_\sigma
> 0$. Then, this $RG$ equation shows an unstable fixed point
$g_{\sigma}^{*} = {\eta_{\sigma}}/{A}$ consistent with our
previous study\cite{Disorder_ASL}. In the case of weak randomness
$g_{\sigma} < g_{\sigma}^{*}$ the $g_{\sigma}$ goes to zero owing
to the first term, implying the emergence of a delocalized spinon
state. In the case of strong randomness $g_{\sigma} >
g_{\sigma}^{*}$ the $g_{\sigma}$ goes to infinity owing to the
second term, indicating localization of the spinons. Solving the
$RG$ equation (9), we obtain the spinon conductance \bqa &&
\sigma_{sn} =
\sigma_{sn}^{0}\Bigl(\frac{L}{l_{e}}\Bigr)^{\eta_{\sigma}}  +
\frac{A}{\eta_{\sigma}}\Bigl[1 -
\Bigl(\frac{L}{l_{e}}\Bigr)^{\eta_{\sigma}} \Bigr] , \eqa where
$\sigma_{sn}^{0} \sim {g_{\sigma}^{0}}^{-1}$ is the bare spinon
number conductance and $l_e$, the elastic mean free
path\cite{Fisher_disorder}. The key feature is the power law
correction resulting from the anomalous critical exponent
$\eta_\psi$. This is in contrast to the free Dirac theory. In the
limit $\eta_{\sigma} \rightarrow 0$ the present conductance
formula is reduced to the logarithmic suppression, $\sigma_{sn} =
\sigma_{sn}^{0} - Aln\Bigl(\frac{L}{l_e}\Bigr)$ consistent with
that of the free Dirac theory\cite{Fisher_disorder}. The density
of the spinon states can be easily calculated in the delocalized
regime. Although it is not clearly defined in the $ASL$, the
terminology is used below as it is. The formal expression of the
density of states is given by $\rho = \lim_{M\rightarrow 0}
\frac{\rho_0}{16M}<Tr[{\bf U}^{\dagger} + {\bf
U}]>$\cite{Fisher_disorder}. We find that the density of states
also has the power law correction \bqa \frac{\rho-\rho_0}{\rho_0}
= -B{g}_{\sigma}\frac{1}{\eta_{\sigma}}(l_{e}^{-\eta_\sigma} -
L^{-\eta_\sigma}) , \eqa where $B$ is a positive numerical
constant. This expression shows a finite density of spinon states
in the limit of $L\rightarrow \infty$, given by $\rho = \rho_{0}(1
- B{g}_{\sigma}\eta_{\sigma}^{-1}l_{e}^{-\eta_\sigma} )$. In the
limit $\eta_{\sigma} \rightarrow 0$ we also recover the
logarithmic suppression $\frac{\rho-\rho_0}{\rho_0} =
-B{g}_{\sigma}ln\Bigl(\frac{L}{l_e}\Bigr)$ in the free Dirac
theory\cite{Fisher_disorder}. On the other hand, in the case of a
negative exponent $\eta_{\sigma} < 0$ the $RG$ equation (9) shows
a runaway characteristic for the $g_\sigma$. This implies the
Anderson localization of spinons and thus the $ASL$ disappears in
the presence of nonmagnetic impurities.

Next we discuss how the delocalized state of spinons can emerge.
In two spacial dimensions the $NL{\sigma}M$ based on the free
Dirac theory does not have a stable fixed
point\cite{Fisher_disorder} indicating a delocalized state. But,
the presence of a topological term such as a Wess-Zumino-Witten
($WZW$) term or Berry phase ($\theta$) term can result in a stable
critical point\cite{WZW}. This is well known in the disordered
metal and antiferromagnetic spin chain with spin $1/2$. On the
other hand, the delocalization in the present study has nothing to
do with such topological terms. It originates from the criticality
of the $ASL$. It is well known that the $NL{\sigma}M$ has an
unstable fixed point above two dimensions\cite{Fisher_disorder}.
The criticality of $\eta_{\psi}
> 0$ leads the $ASL$ to be in $(2+2\eta_{\psi})$ dimensions
effectively. As a result the delocalization emerges against weak
randomness. We would like to stress that the mechanism of
delocalization in the critical phase ($ASL$) totally differs from
that arising from the topological terms. This is reflected in the
$RG$ equation. In the delocalization induced by the topological
terms the stable fixed point appears from the $g_{\sigma}^{2}$
term in the $RG$ equation\cite{WZW}. This is associated with
destructive interference effect of interactions. On the other
hand, the delocalization driven by the criticality arises from the
$g_{\sigma}$ term [Eq. (9)]. This is due to increase of effective
dimensionality, as discussed above.

If the $WZW$ term appears in the $NL{\sigma}M$ Eq. (8), a stable
strong coupling fixed point is expected to exist\cite{WZW} in both
cases, $\eta_{\psi} > 0$ and $\eta_{\psi}< 0$. The density of
states would exhibit a power law behavior near the fixed point.
Moreover, it is expected to vanish as energy decreases down to
zero\cite{WZW}. This is in contrast to the present case. However,
it is not clear whether this new fixed point is stable against
instanton excitations. The spinons would be renormalized by the
disorder effect. It is difficult to determine how the renormalized
spinons affect the internal gauge charge. This problem will be an
important future work on the stability of the $ASL$ against both
instantons and disorder.

Our present study has important implications in the role of
nonmagnetic disorder in high $T_c$ cuprates. According to one
scenario\cite{Doped_PMMI_SC}, the pseudogap ($PG$) state is
proposed to be the $ASL$. Then, the $ASL$ should be stable against
disorder because all samples include disorder. The stability of
the $ASL$ against disorder depends on the sign of the critical
exponent $\eta_\psi$. As pointed out earlier, most
evaluations\cite{Wen_ARPES,Ye_propagator,Khveshchenko_propagator,QED_eta}
support a negative critical exponent. This would destabilize the
$ASL$ against disorder, resulting in the Anderson localization. A
positive critical exponent\cite{Tesanovic_QED} is necessary for
the stability of the $ASL$. Our investigation requires more
careful determination of the critical exponent $\eta_{\psi}$.
Furthermore, the present study tells us that the previous studies
of disorder effects in the $PG$ state\cite{Nagaosa_disorder} are
difficult to be applied. This is because the
studies\cite{Nagaosa_disorder} are based on the free fermion
theory ignoring strong gauge fluctuations. The existence of the
delocalized state in the impurity doped $ASL$ seems to be
inconsistent with experiments\cite{Nagaosa_disorder}. In
experiments a nonmagnetic impurity is believed to localize spin
$1/2$. This trapped spin acts as a free spin, showing the
Curie-Weiss behavior in the spin
susceptibility\cite{Nagaosa_disorder}. We expect that this
inconsistency may be resolved by considering strong disorder. We
note that strong disorder causes the localization of spinons. $Zn$
impurity can be considered to be a strong scatterer owing to its
compact electronic shell structure.

In the present study we did not consider the effect of hole
doping. Doped holes are represented by holons in the context of
the slave boson theory. Holons can affect gauge fluctuations. For
example, when holons are condensed, gauge fluctuations become
massive via the Anderson-Higgs mechanism. As a result a free Dirac
theory is obtained. The spinon can be localized even in weak
disorder. The Curie-Weiss behavior can be easily understood based
on the free Dirac theory in the superconducting
state\cite{Free_Dirac_disorder_Kim}. In the $PG$ phase the holons
are not condensed. Gauge fluctuations remain massless. In this
case damping effect in gauge fluctuations is expected to arise
from holon contributions\cite{Damping,Kim_damping}. The role of
disorder in the presence of damped gauge fluctuations would be an
interesting future work\cite{Kim_damping}.

Lastly, we would like to comment about the present approach based
on the $NL\sigma{M}$. The $NL{\sigma}M$ approach has some
advantages, compared with the previous access based on the
fermionic action\cite{Disorder_ASL}. First, the $NL{\sigma}M$
approach is more efficient than the fermionic one in investigating
physics of phase transitions. The $NL{\sigma}M$ assumes the
presence of a local order parameter. On the other hand, the
previous study\cite{Disorder_ASL} is based on the absence of local
ordering. If we start from a disordered phase without local
ordering and approach a critical point associated with a phase
transition, a local order parameter is to emerge, strongly
fluctuating. In order to describe critical fluctuations of the
order parameter, we explicitly introduce the local order parameter
and obtain an effective field theory of the order parameter field
by integrating over the fermions. The resulting effective field
theory is the Ginzburg-Landau free energy. As a matter of fact the
$NL{\sigma}M$ is nothing but the Ginzburg-Landau free energy
formulation. If we start from the fermion action, we realize that
it is not easy to describe phase transitions. This is because
phase transitions are basically nonperturbative phenomena not
captured by perturbative calculations based on the fermionic
action. In order to describe the phase transitions it is necessary
to sum infinitely many diagrams in the fermionic action. On the
other hand, in the Ginzburg-Landau formulation the phase
transitions are easily described. This is the reason why the
$NL\sigma{M}$ is called an effective field theory for phase
transitions. Second, it is also easy to calculate some physical
quantities like conductance and density of states. These
quantities directly appear in the $NL\sigma{M}$. We can easily
obtain the conductance and density of states as a function of
energy and size of a system by the present $RG$ calculation. Last,
topologically nontrivial excitations in the $NL\sigma{M}$ are not
captured in the fermionic action although these are not
intensively discussed in the present paper.

In summary, we examined the effect of nonmagnetic disorder on the
$ASL$ criticality characterized by its anomalous critical
exponent. In the case of a positive exponent the critical point
remains stable at least against weak disorder. But, in the
opposite case the critical point becomes unstable.


\begin{thebibliography}{9}
\bibitem{Nc} The total flavor number $N_t$ of massless fluctuations is
given by $N_t = N + N_b$. Here $N$ is the flavor number of
massless Dirac fermions and $N_b$, that of critical holon
fluctuations originating from doped holes. In the SU(2) slave
boson theory\cite{Doped_PMMI_SC} $N_b$ is given by $2$. Thus, the
total flavor number is $N_t = 4$. This exceeds the usually chosen
critical value $N_c \approx 3.24$\cite{DonKim_QED}. The
$S\chi{S}B$ is not expected to occur.
\bibitem{Saddle_point} An explicit integration results in the following expression
$\frac{\pi}{2\Lambda^2}\frac{Q^2}{W} =
Hypergeometric2F1[\frac{1}{1 - \eta_{\psi}}, 1, 1 + \frac{1}{1 -
\eta_{\psi}}, -\frac{(\Lambda^2)^{1-\eta_{\psi}}}{Q^2}]$. This is
done by Mathematica 5.0.
\end{thebibliography}
\end{document}